\documentclass{elsarticle}
\usepackage{hyperref}
\usepackage{graphicx}  
\usepackage{dcolumn}   
\usepackage{bm}        
\usepackage{amssymb}   
\usepackage{amsfonts}
\usepackage{graphics}
\usepackage{grffile}   
\usepackage{epsfig}
\usepackage{multirow}
\usepackage{wrapfig}
\usepackage[usenames]{color}
\usepackage[normalem]{ulem} 
\usepackage{color}
\usepackage[utf8]{inputenc}
\usepackage[T1]{fontenc}
\usepackage{amsmath}
\usepackage{ulem}
\usepackage{array}
\usepackage{subcaption}
\usepackage{siunitx}
\DeclareSIUnit\clight{\text{\ensuremath{c}}}
\biboptions{numbers,sort&compress} 
\usepackage[symbol*]{footmisc}

\DeclareSIUnit\ele{e^{\text{-}}}

\newcolumntype{P}[1]{>{\centering\arraybackslash}p{#1}}
\newcolumntype{M}[1]{>{\centering\arraybackslash}m{#1}}

\makeatletter
\def\ps@pprintTitle{%
  \let\@oddhead\@empty
  \let\@evenhead\@empty
  \def\@oddfoot{\reset@font\hfil\thepage\hfil}
  \let\@evenfoot\@oddfoot
}
\makeatother

\begin{document}
\begin{frontmatter}
\title{Detection efficiency and spatial resolution of Monolithic Active Pixel Sensors bent to different radii}

\author[a]{Anton Andronic}
\author[b]{Pascal Becht}
\author[b,c]{Mihail Bogdan Blidaru}
\author[d,e]{Giuseppe Eugenio Bruno}
\author[f]{Francesca Carnesecchi}
\author[g]{Emma Chizzali}
\author[d,e]{Domenico Colella}
\author[h,i]{Manuel Colocci}
\author[j,k]{Giacomo Contin}
\author[g]{Laura Fabbietti}
\author[g]{Roman Gernh\"auser}
\author[f]{Hartmut Hillemanns}
\author[h,i]{Nicolo Jacazio}
\author[f]{Alexander Philipp Kalweit}
\author[f]{Alex Kluge}
\author[l]{Artem Kotliarov}
\author[l]{Filip Křížek}
\author[f,g]{Lukas Lautner\corref{cor1}}
\ead{lukas.lautner@tum.de}
\cortext[cor1]{Corresponding author}
\author[f]{Magnus Mager}
\author[f]{Paolo Martinengo}
\author[b]{Silvia Masciocchi}
\author[b,f]{Marius Wilm Menzel}
\author[m]{Alice Mulliri}
\author[f]{Felix Reidt}
\author[n,o]{Riccardo Ricci}
\author[p]{Roberto Russo}
\author[b]{David Schledewitz}
\author[h,i]{Gilda Scioli}
\author[q]{Serhiy Senyukov}
\author[m]{Sabyasachi Siddhanta}
\author[p]{Jory Sonneveld}
\author[b]{Johanna Stachel}
\author[f]{Miljenko Šuljić}
\author[a,f]{Nicolas Tiltmann}
\author[d,e]{Arianna Grisel Torres Ramos}
\author[g]{Berkin Ulukutlu}
\author[m,s]{Gianluca Usai}
\author[f]{Jacob Bastiaan Van Beelen}
\author[q,t]{Yitao Wu}
\author[b]{Alperen Y\"unc\"u}

\address[a]{Universit\"at M\"unster, M\"unster, Germany}
\address[b]{Ruprecht Karls Universit\"at Heidelberg, Heidelberg, Germany}
\address[c]{GSI Helmholtzzentrum für Schwerionenforschung GmbH, Darmstadt, Germany}
\address[d]{Dipartimento di Fisica “M. Merlin” dell’Università e del Politecnico di Bari, Bari, Italy}
\address[e]{INFN, Sezione di Bari, Bari, Italy }
\address[f]{European Organisation for Nuclear Research (CERN), Geneva, Switzerland}
\address[g]{Technische Universit\"at M\"unchen, M\"unchen, Germany}
\address[h]{INFN, Sezione di Bologna,Bologna, Italy}
\address[i]{Dipartimento Fisica e Astronomia, Università di Bologna,Bologna, Italy}
\address[j]{INFN, Sezione di Trieste, Trieste, Italy}
\address[k]{Università degli studi di Trieste, Trieste, Italy}
\address[l]{Nuclear Physics Institute of the Czech Academy of Sciences, Husinec- Rez, Czechia}
\address[m]{INFN, Sezione di Cagliari, Cagliari, Italy}
\address[n]{INFN,Sezione di Salerno, Salerno, Italy}
\address[o]{Università degli Studi di Salerno, Salerno, Italy}
\address[p]{Nikhef National institute for subatomic physics, Amsterdam, Netherlands}
\address[q]{Université de Strasbourg, CNRS, IPHC UMR 7178, Strasbourg, France}
\address[r]{Università degli studi di Cagliari, Cagliari, Italy}
\address[s]{University of Science and Technology of China, Hefei, China}



\begin{abstract}

Bent monolithic active pixel sensors are the basis for the planned fully cylindrical ultra low material budget tracking detector ITS3 of the ALICE experiment.
This paper presents results from testbeam campaigns using high-energy particles to verify the performance of \SI{50}{\um}~thick bent ALPIDE chips in terms of efficiency and spatial resolution. The sensors were bent to radii of 18, 24 and \SI{30}{\mm}, slightly smaller than the foreseen bending radii of the future ALICE ITS3 layers. 
An efficiency larger than $99.9\%$ and a spatial resolution of approximately \SI{5}{\um}, in line with the nominal operation of flat ALPIDE sensors, is obtained at nominal operating conditions. These values are found to be independent of the bending radius and thus constitute an additional milestone in the demonstration of the feasibility of the planned ITS3 detector. 
In addition, a special geometry in which the beam particles graze the chip and traverse it laterally over distances of up to \SI{3}{\mm} is investigated.

\end{abstract}


\end{frontmatter}

\section{Introduction}

Over the last decades, silicon-based detectors have become the main tracking devices close to the interaction point in many high-energy and nuclear physics experiments.
In order to minimise the distance between the sensors and the primary vertex, and to reduce the overall material budget of the detector, truly cylindrical, bent, wafer-size Monolithic Active Pixel Sensors (MAPS) are employed.
These sensors form the basis of the ITS3 (Inner Tracking System 3)~\cite{loi,TDRITS3} upgrade of the vertex detector of the ALICE collaboration, which is scheduled for installation during Long Shutdown 3 (2026--2029) at the LHC.
For the ITS3, it is planned to employ large area sensors that are bent to half-cylinders with radii of 19, 25.2 and~\SI{31.5}{\mm}. In this context, the performance of MAPS in terms of detection efficiency and space point resolution for small bending radii needs to be assessed.
The studies presented in this article constitute an important R\&D step with respect to the first ever performance study of bent monolithic active pixel sensors in which it was shown that the bent chips preserve their full electrical functionality and particle detection capability.
The results shown here complement and extend the prior findings~\cite{ALICEITSproject:2021kcd} by investigating the three radii compatible with the values foreseen for the ITS3 project as well as by studying a different bending axis and a setup in which also the periphery of the chip is bent. This was possible thanks to a new assembly method which leaves no tape or other material in the relevant region in front or behind the chip. 
Following up on the track to hit residuals presented in Ref.~\cite{ALICEITSproject:2021kcd}, the actual spatial resolution for bent MAPS as a function of threshold is determined for the first time. For reference, the findings for the bent chips are directly compared to those of flat chips obtained with similar setups. 
The experimental setup used to study the bent sensors also allows for an unique geometry where the beam particles graze the chip. In this configuration, which is studied for the first time in ALPIDEs, the particles traverse the sensitive area of the chip over distances of several millimeters.
\section{The bent ALPIDE chip}
\label{sec.:BentChip}

\begin{figure}[thp]
	\centering
    \includegraphics[width=\textwidth]{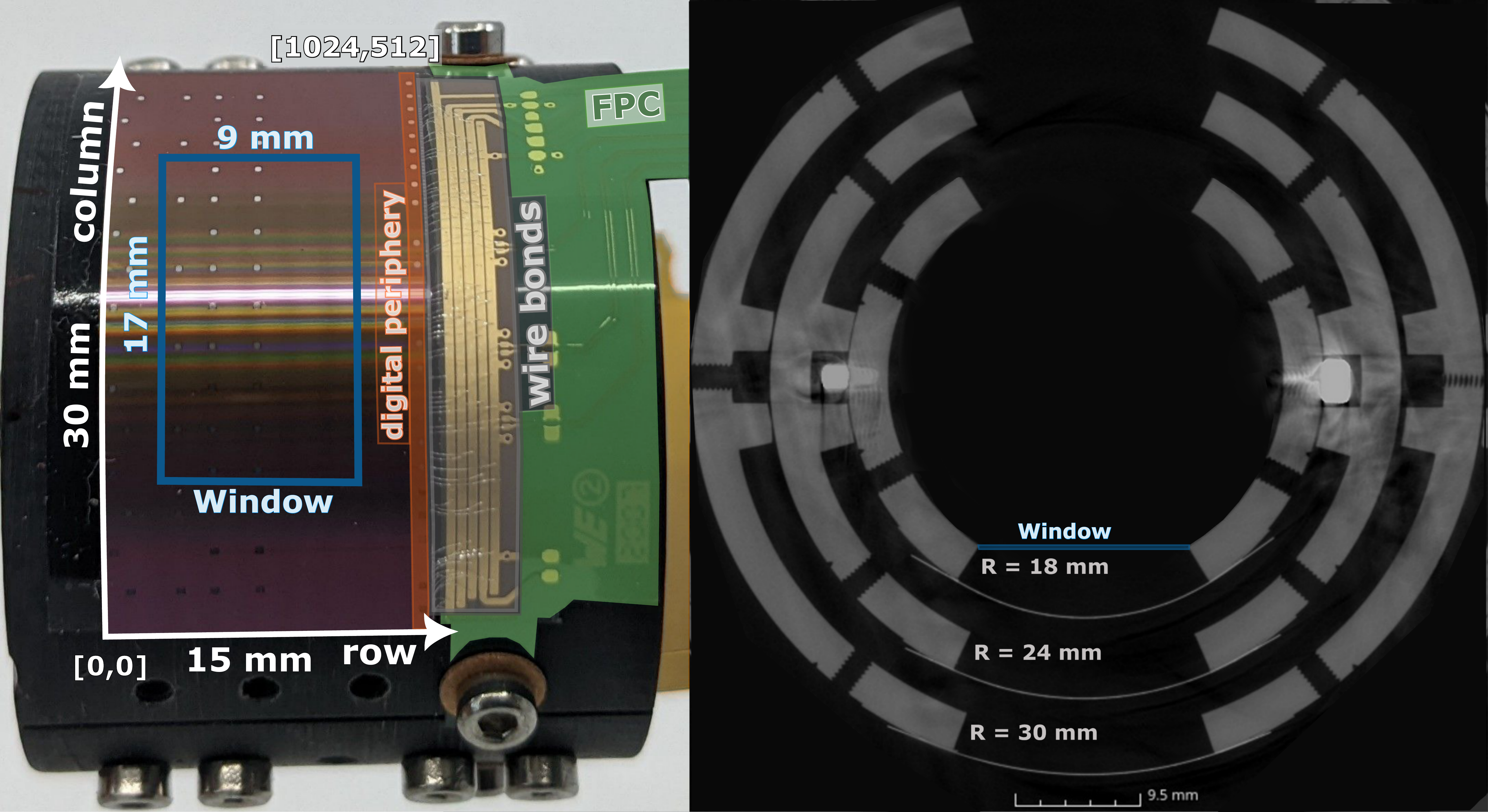}    

	\caption{(Left) Close-up of an ALPIDE sensor bent to a radius of \SI{18}{\mm} fixed onto the innermost part of a cylindrical jig. The pads on the digital periphery of the chip are used to bond to a FPC which connects to the readout electronics. The opening in the jig behind the sensor is indicated. (Right) Computer tomographic scan of a similar jig with three bent ALPIDE chips visible as faint lines.}
	\label{fig:jigFourbent}
\end{figure}

As in the previous study, readily available ALPIDE chips are used in the present work. The ALPIDE sensor was developed by the ALICE Collaboration for its first upgrade of the Inner Tracking System~(ITS2) \cite{TDR,ALPIDE-proceedings-1,ALPIDE-proceedings-2,ALPIDE-proceedings-3}. The chip measures \qtyproduct{30x15}{\milli\meter} and is produced in the TowerJazz \SI{180}{\nm}~CMOS imaging process~\cite{towerjazz}.
For this study, chips thinned down to \SI{50}{\um} that have a \SI{25}{\um}-thick epitaxial layer were used. 
The sensor features a matrix of \qtyproduct{1024x512} (column~$\times$~row) pixels with binary output.
The pixels of size \qtyproduct{29.24x26.88}{\micro\meter} contain the sensing diode connected to its individual and continuously active front-end amplifier, shaper, discriminator and multiple-event buffer, which is not continuously active. They also contain analog and digital testing circuitry, e.g. for adjusting the charge threshold of the pixel by injecting a programmable test charge into the sensing node. The charge threshold can be changed for all pixels simultaneously by adjusting the front-end parameters~\cite{ALPIDE-proceedings-1, ALPIDE-proceedings-2, ALPIDE-proceedings-3, pulselength}. 
The priority encoding circuits, located in-between the column pairs, propagate the addresses of the hit pixels to the digital circuitry on the chip periphery. The digital periphery occupies an area of \qtyproduct{1.2x30}{\milli\meter} along the edge of the chip.
A series of aluminum pads at the periphery are used to electrically interface to the sensor via wire bonds (see Fig.~\ref{fig:jigFourbent}). 

A special procedure to bend and simultaneously fix the chip on a 3D printed cylindrical jig was applied. Compared to the results presented in~\cite{ALICEITSproject:2021kcd} where the sensor was wire-bonded before being bent, this study presents a new procedure where the sensors are bonded after being bent and placed on a cylindrical jig.
This became the current baseline for future prototypes and final detector implementation. In order for this to be possible the electrical substrate, to which the readout system interfaces, needs to be bent to the same radius as the chip. Therefore, an FPC (flexible printed circuit) is used that ensures the electrical connection between the bent sensor and the readout electronics. 
The same bending procedure was applied independently for all chips and jig radii. Sensors attached in this way to a 3D printed cylinder are shown in Fig.~\ref{fig:jigFourbent}. Each jig has a wall thickness of \SI{3.1}{\mm} and is made of polypropylene-like \mbox{ACURRA-25} material. It constitutes a significant amount of material that leads to additional scattering. 
Therefore, a window of \qtyproduct{17x9}{\milli\meter\squared} is cut out of the jig behind the chip to minimise the material crossed by the beam particles.

For bending and attaching the chip, two different techniques were used.  In both cases, first, the FPC is fixed on the jig by two screws. Afterwards, in one technique (procedure A), a bi-adhesive tape is applied to the jig and the no-stick cover is peeled off. The part of the tape covering the cut-out window of the jig is then removed, and the chip is placed on top. It is slowly bent by rotating the jig with a stepping motor, while being temporarily held in place by a foil under tension.
In the other technique (procedure B), a foil with a central cut-out larger than the window in the jig is fixed to the jig on one side of the chip by three screws. The sensor is held along half its length by a vacuum device and placed between the foil and jig. The jig is then slowly rotated, while holding the foil under tension, bending the sensor. Finally, the other end of the foil is fixed again with three screws to the jig keeping the chip bent without any glueing. We accommodate different radii within a single assembly by nesting the individual cylindrical jigs inside one another.

\begin{figure}
	\centering
	 \includegraphics[width=0.9\textwidth]{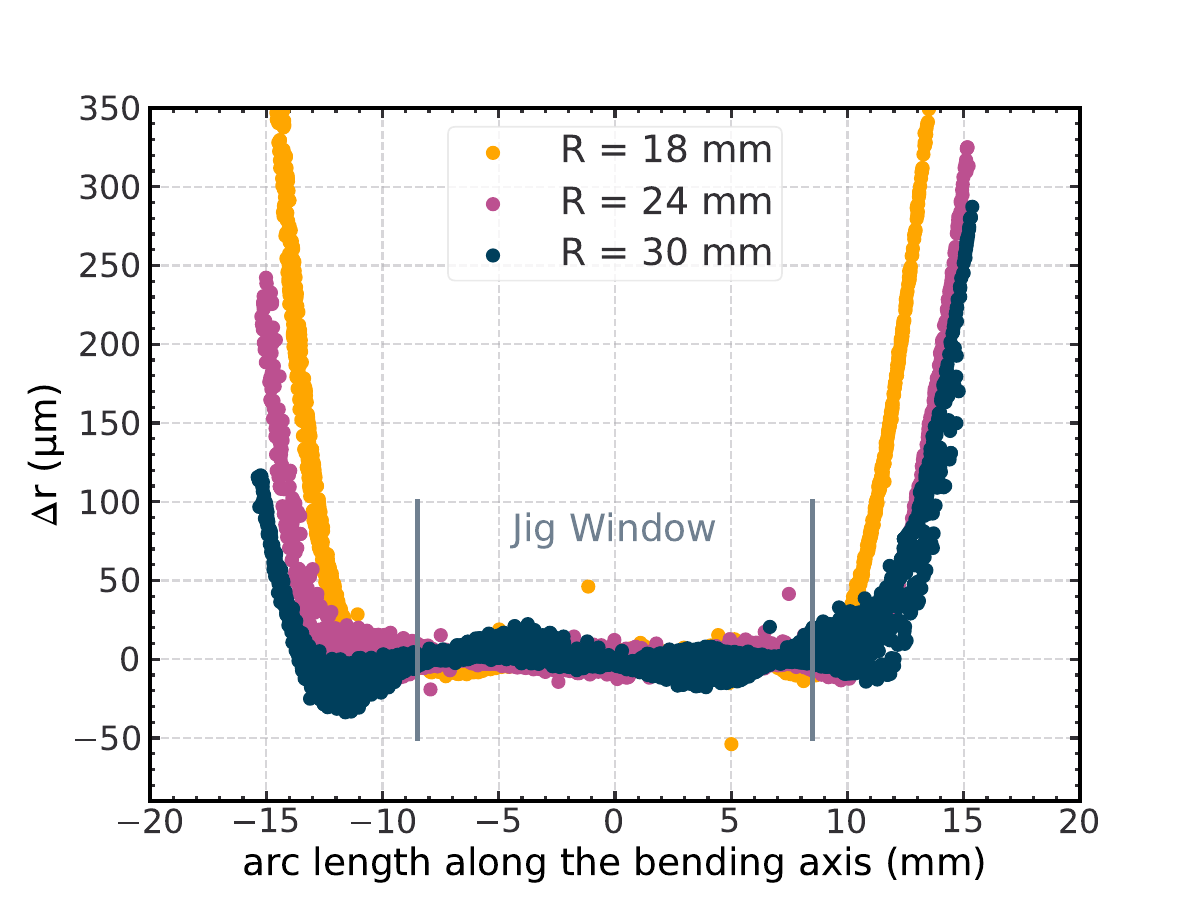}
	\caption{Deviation in the radial direction ($\Delta r$) from a perfect cylindrical geometry as a function of the arc length along the bending axis. Results are shown for the chips bent with the procedure A (see text for details), as measured with an optical 3D profilometer.}
	\label{fig:Profilometer} 
\end{figure}

The precision of the mounting procedure and assembly of the bent sensors was tested with a 3D optical profilometer. The assumption of a perfect cylindrical geometry was verified to be correct within \SI{50}{\um} around an area of \SI{\pm 10}{\mm} from the center, as shown in Fig.~\ref{fig:Profilometer}. 
A detachment of the sensor edges along the bending axis can be observed in procedure A.
\section{Testbeam setup}
\label{sec:setup}

The results presented in this document are based on data collected during several testbeam campaigns at the DESY II testbeam facility~\cite{desyII} (December 2019, June 2020, April 2021) and the CERN SPS (July 2021).
At DESY, the beam line 24 with a \SI[per-mode=symbol]{5.4}{\giga\electronvolt\per\clight} electron beam was used. At SPS, the H6 beam line provided \SI[per-mode=symbol]{120}{\giga\electronvolt\per\clight} pions, protons and muons and electrons with a rough relative admixture of 60--70\%/25\%/5--15\%. In both cases, the energy loss of the beam particles in silicon is close to the values of a minimum ionizing particle.
The setup with bent ALPIDEs at DESY(SPS) consisted of a telescope with \num{4}(\num{6})~flat ALPIDE chips used as reference planes, \num{2}(\num{3}) on each side of the Devices-Under-Test (DUTs). 
The data for flat ALPIDEs as the DUT were taken at DESY and used a telescope with \num{6}~reference planes.

\begin{figure}[thp]
	\centering
    	\includegraphics[width=0.95\textwidth]{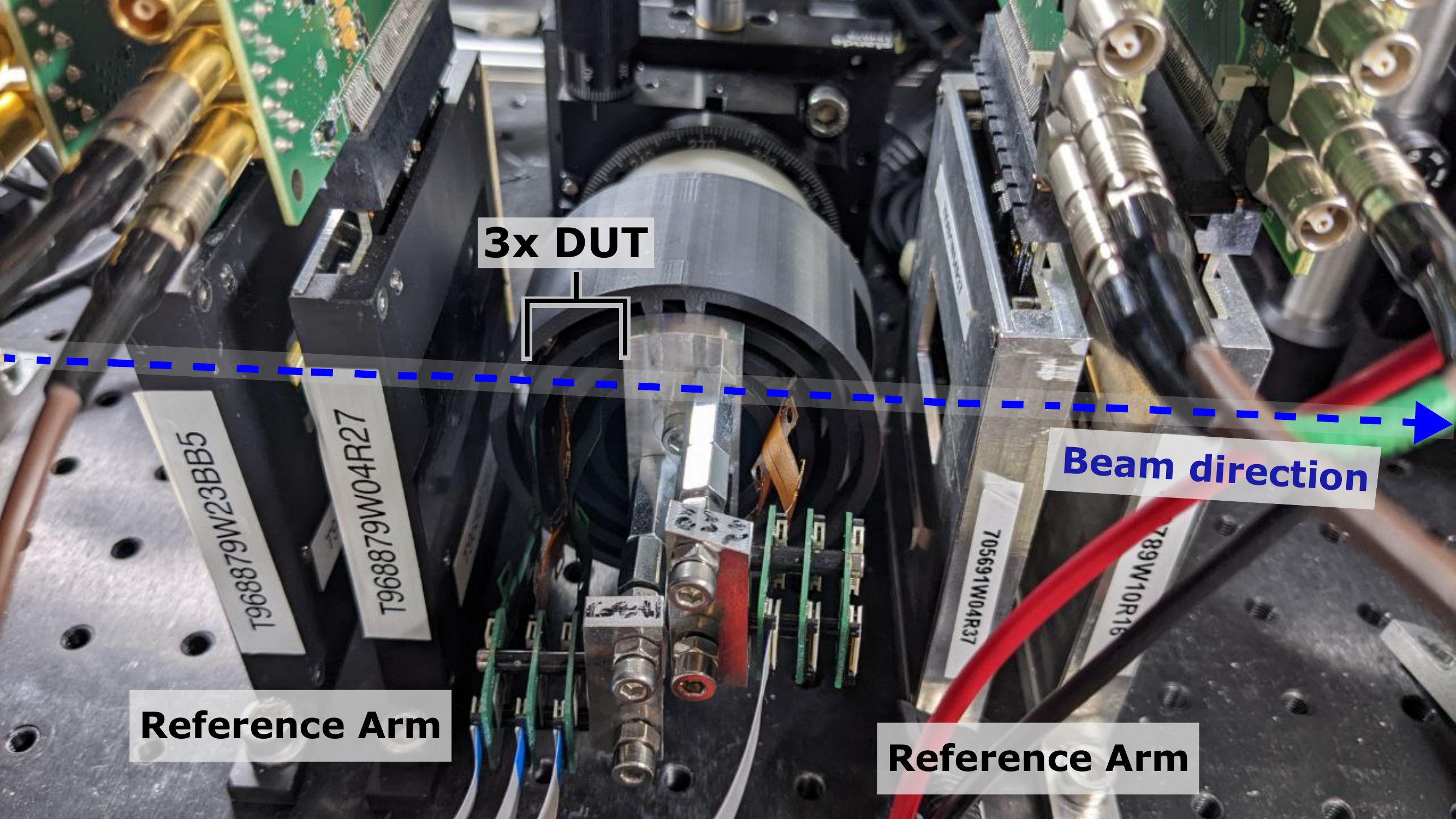}
\includegraphics[width=0.95\textwidth]{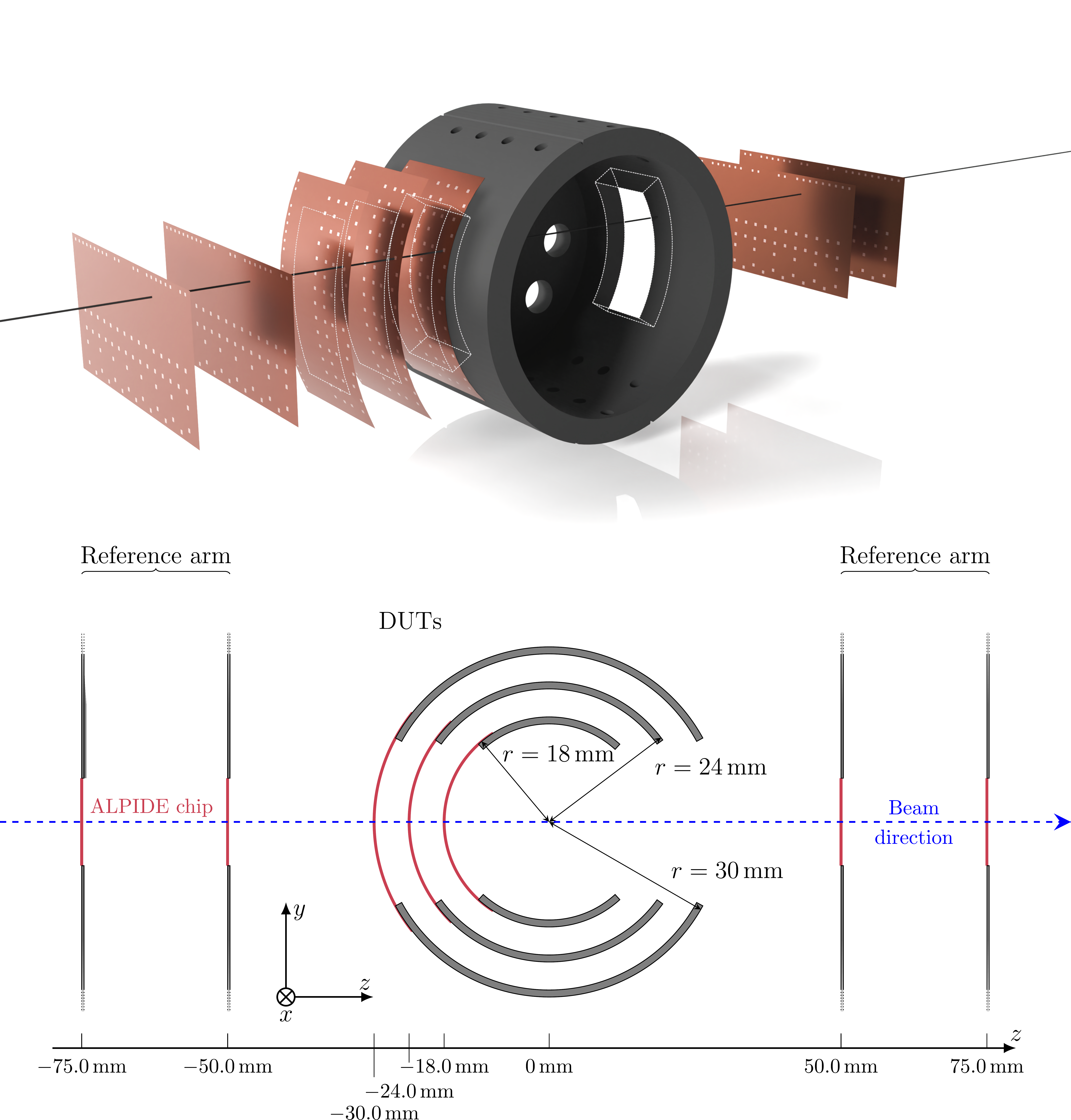}

\caption{ (Top) Photograph of the multi-DUT setup;
(Center) 3D rendering of the setup showing 4 reference planes and 3 bent DUTs. Only the jig of the innermost bent sensor is shown. (Bottom) Schematic drawing of the geometry of the setup.}
	\label{fig:setup_sketch}
\end{figure}

For all the setups, the reference planes were operated at \SI{-3}{\volt} reverse substrate bias, whereas the DUTs were operated without reverse substrate bias. An overview of the testbeam campaigns is given in Tab.~\ref{tab:testbeams}. The setups used in all the campaigns are very similar. In the following lines, the April 2021 setup is described in detail. The main difference with respect to the setups used in the other campaigns is given by the number of reference planes and the geometry of the DUT (flat, bent, or grazing). In addition, the bending procedure A was applied for the April 2021 testbeam and the procedure B for the July 2021 testbeam. The so-called grazing beam geometry is described in Sec.~\ref{sec.:grazing}.
The DUTs are bent to the ITS3 radii as originally laid out in the letter of intent~\cite{loi} and placed in the middle of the setup, as shown in Fig.~\ref{fig:setup_sketch}.
This placement causes the column and row axes of the DUT to be swapped with respect to the reference planes.
The flat DUTs were also put in the middle of the setup with a \SI{2.5}{\centi\meter} spacing between all planes, but in the same orientation as the reference planes.

\begin{table}[]
\begin{center}

\begin{tabular}{M{0.2\linewidth }M{0.2\linewidth} M{0.2\linewidth}M{0.2\linewidth}}

\hline \hline
    DUT   &  Date &  Facility & Number of reference planes \\ \hline  \hline
\multirow{2}{*}{Flat} &  December 2019 & DESY & 6 \\
                  & June 2020 & DESY &  6\\ \hline
\multirow{2}{*}{Bent} & April 2021 &  DESY & 4 \\
                  & July 2021 & SPS & 6 \\\hline
Grazing & December 2020 & DESY & 6 \\\hline
\end{tabular}
\caption{Summary of the test beam campaigns.} \label{tab:testbeams}
\end{center}
\end{table}

The standard readout system of the ALPIDE~\cite{9508095} is used with a small adapter board attached to each FPC.

This board serves as interface between the FPC and the PCIe connector of the readout electronics. It primarily contains circuitry needed for voltage regulators, and active level shifters and filter capacitors to minimize the impact of transient currents and filter power supply noise. The bias voltage is decoupled via a pi low-pass filter circuit. 
The best performance was achieved with
\SI{10}{\micro\farad} filter capacitance and allow operation at low thresholds (< 90 $e^{-}$) and low noise levels ($\ll$ 10 noisy pixels per sensor).

The trigger was given by the coincidence of the discriminated output of two photomultiplier tubes operated in the plateau region connected to two scintillators placed in front and behind of the telescope. The scintillators had a size of \qtyproduct{4.5x2.5}{\cm\squared}, an area slightly larger than the surface of the reference planes.
The triggering logic includes an event separation time of \SI{100}{\micro\second} and a past protection time of \SI{20}{\micro\second}, i.e.~a veto on triggering in the \SI{20}{\micro\second} window following a scintillator output signal. The latter is added to avoid pile-up, as the ALPIDE in-pixel amplifier pulse can reach lengths of a few~tens~of~\si{\us} for very low threshold values \cite{pulselength}.
The data acquisition was based on the EUDAQ2 framework~\cite{eudaq}. 
The results presented in this study contain data from more than 100 runs of at least 100k events, each run corresponding to different thresholds and DUT positions.

\section{Analysis tools and methods} \label{sec.:analysistool}

Data were processed in the Corryvreckan reconstruction framework~\cite{Dannheim_2021}.
Events are built from hits belonging to the same trigger ID on all planes.
Noisy pixels with a firing rate higher than 1000 times the average of the whole chip are masked and not considered in the analysis. This resulted in masking of up to 5 pixels per plane. Adjacent firing pixels per event are grouped together into clusters. The cluster position is calculated as the center-of-gravity of the hits.

A prealignment is achieved by a shift along the global $x$ and $y$ coordinates given by the correlation of the spatial position of the clusters between the planes. In order to select only particles which crossed the least amount of material, a region of interest (ROI) inside the jig window was defined, corresponding to \qtyproduct{14.16 x 8.04}{\mm\squared}. Track candidates are reconstructed by fitting clusters in the reference planes using a general broken lines~\cite{BLOBEL20111760} approach.
The material of all chips (reference planes, DUT, non-DUT bent chips) is considered in the tracking procedure. 
Only tracks which have at least one cluster on each of the reference planes are kept.
In addition, events with more than one reconstructed track are discarded.
The initial parameters for alignment were determined by minimizing the profile of the mean and the RMS of the spatial residuals across the surface of the chip by varying in coarse steps the orientation and position of the DUT.

The values for the found minimum are then fed as starting values to the Millepede-II algorithm implementation in Corryvreckan in order to perform a simultaneous fit of all tracks and determine the final alignment corrections. 
Only the tracks that have a $\chi^{2}/n_{\rm  dof} < 3$ are considered for subsequent analysis, assuming a constant position resolution ($\sigma_{x},\sigma_{y} \sim \SI{5}{\micro\meter}$) for each reference plane.
The reconstructed tracks are then propagated to the DUTs to evaluate the efficiency and resolution of the bent sensors.
On each DUT a circular search window with a radius of \SI{100}{\um} is used around the extrapolated track intercept on the chip surface.
Clusters on the DUT are matched to the track if they are found to be within the search window.

\begin{figure}[t]
  \centering

\includegraphics[width=\textwidth]{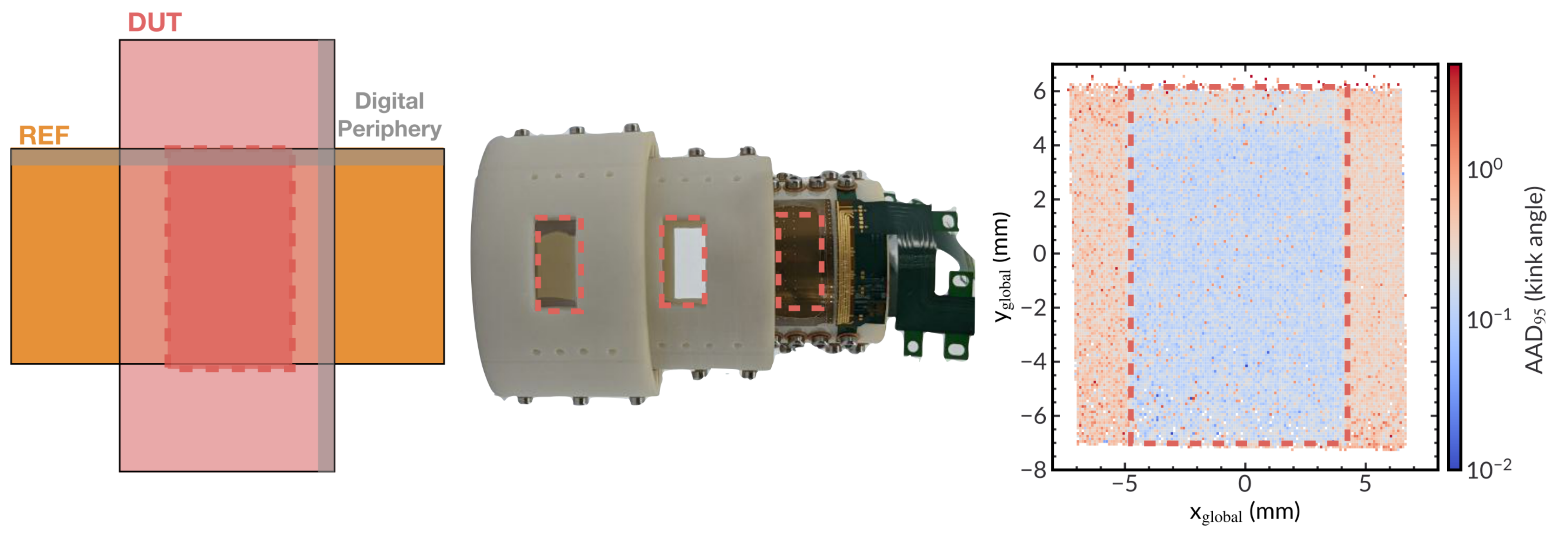}
  \caption {
  (Left) The region of interest (ROI) corresponds to the area covered simultaneously by the reference planes and the opening in the jig.
  The relevant opening window on the jig behind the chip (centre) is clearly identifiably as a region of reduced multiple scattering and a correspondingly narrower kink angle distribution (right). See text for details. 
  Slightly larger scattering angles close to the top edge of the reference planes is visible as a small red band around \SI[product-units=single]{5}{\mm} < $y$ < \SI[product-units=single]{6}{\mm}. This is due to additional material from the carrier cards of the reference planes which overlap with the upper part of the active area of those planes.
 }
  \label{fig:roi_sketch}
\end{figure}

\begin{figure}[t]
  \centering
 
    \includegraphics[width=0.9\textwidth]{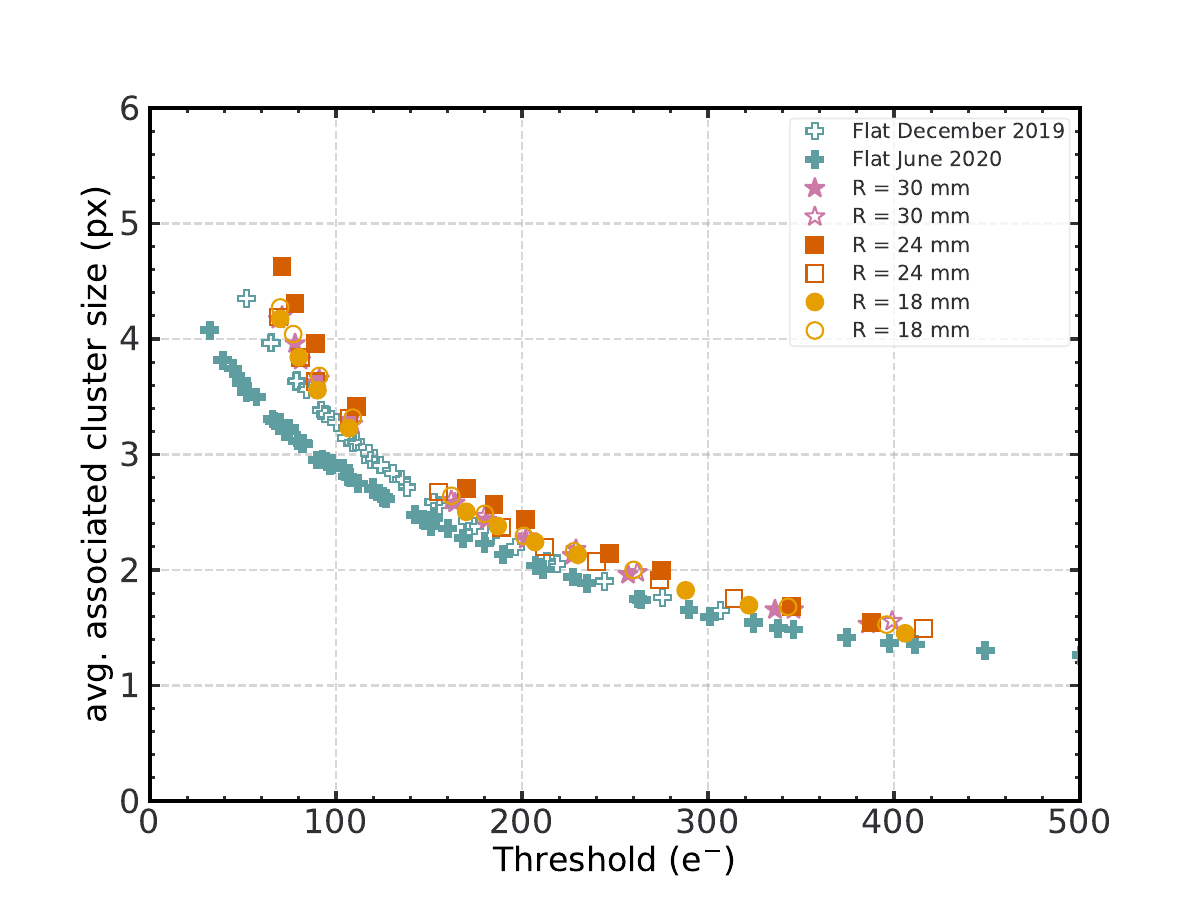}
  \caption {
      Average size of associated clusters as a function of the threshold for different bent and flat ALPIDE chips. The statistical uncertainties are smaller than the marker size. 
  }
  \label{fig:clsize_vs_thr}
\end{figure}

As shown in Fig.~\ref{fig:roi_sketch}, the presence of the jig material produces additional scattering, which manifests itself in a broader kink angle distribution. The latter corresponds to the angle between the two tracklets that are determined using only the upstream and downstream reference planes, respectively. In order to minimise the effect of the tail of the scattering angle distribution, the absolute average deviation of the lower 95\% of the distribution (AAD$_{95}$) is used. The AAD is defined as AAD~$= \frac{1}{n}\sum_{i=1}^{n}|x_{i} - \overline{x}|$, for a sample of points $x_{i}$ with arithmetic mean $\overline{x}$.
As can be seen, tracks passing through the ROI, are not subject to large angle scattering except for a small area close to the top edge of the reference planes. The latter is due to additional material from the carrier cards of the reference planes which overlap with the upper part of the active area of those planes. The ROI region as shown in Fig.~\ref{fig:roi_sketch} was used to determine the spatial resolution and the efficiency. Including or excluding the small area close to the top edge of the reference planes in the analysis did not lead to significant differences in the final results.
In the following sections, results are presented as a function of the threshold. 
The relationship between the average cluster size and threshold is shown in Fig.~\ref{fig:clsize_vs_thr} for two different flat sensors and the six bent senors from the July 2021 testbeam. \newpage
The threshold values shown here were determined by internal pulsing as described in Sec.~\ref{sec.:BentChip}. As expected the cluster size is decreasing with increasing threshold.
No significant difference between flat and bent sensors has been observed\footnote{The slightly larger values for the bent sensors were likely caused by the non-zero incident angle in the bent configuration (up to 10$^\circ$).
Since the standard threshold determination via internal pulsing for the April 2021 test beam was not set up properly, the actual threshold values for this campaign were reconstructed based on the average associated cluster size (see~\ref{sec.:appendixA} for details).}.

\section{Results} \label{sec.:results}

\subsection{Efficiency} 

\begin{figure}[!tbp]
  \centering

    \includegraphics[width=0.9\textwidth]{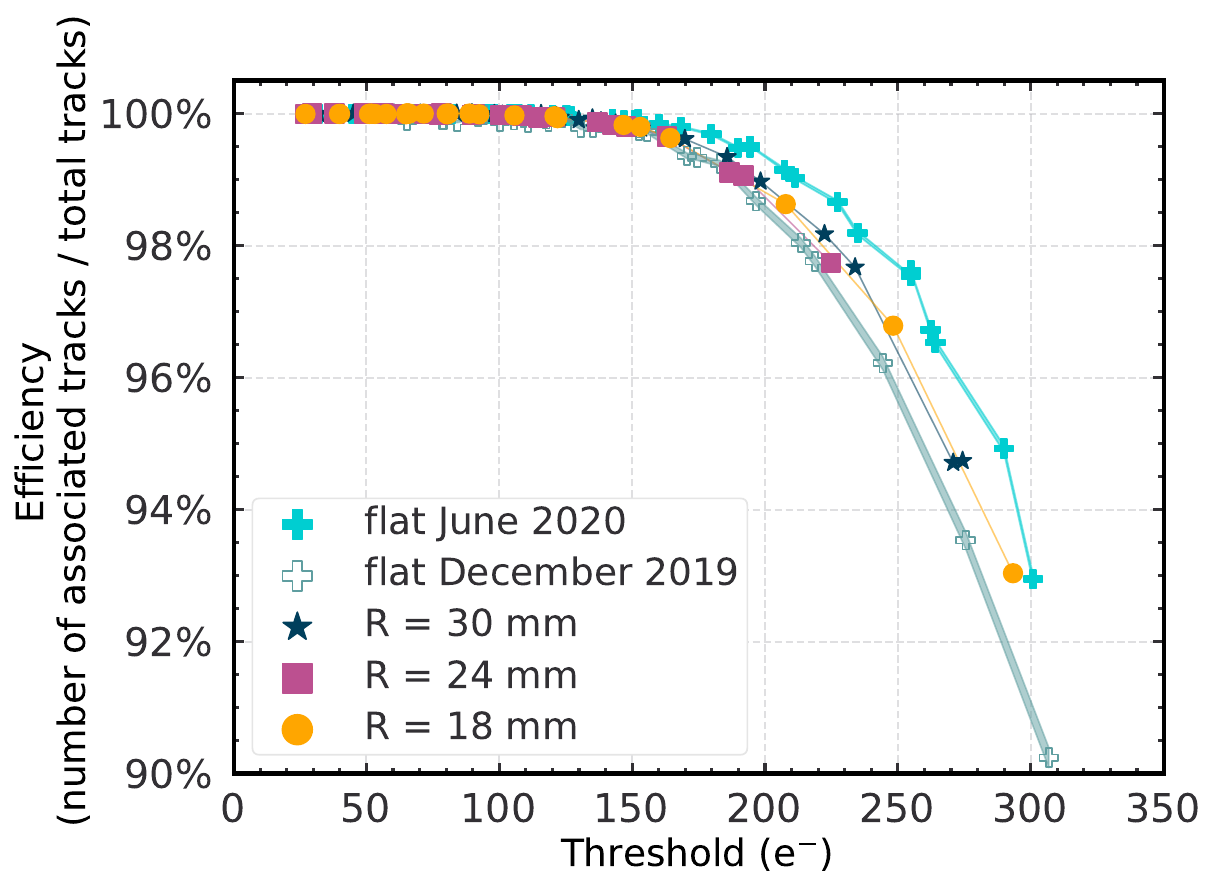}
\includegraphics[width=0.9\textwidth]{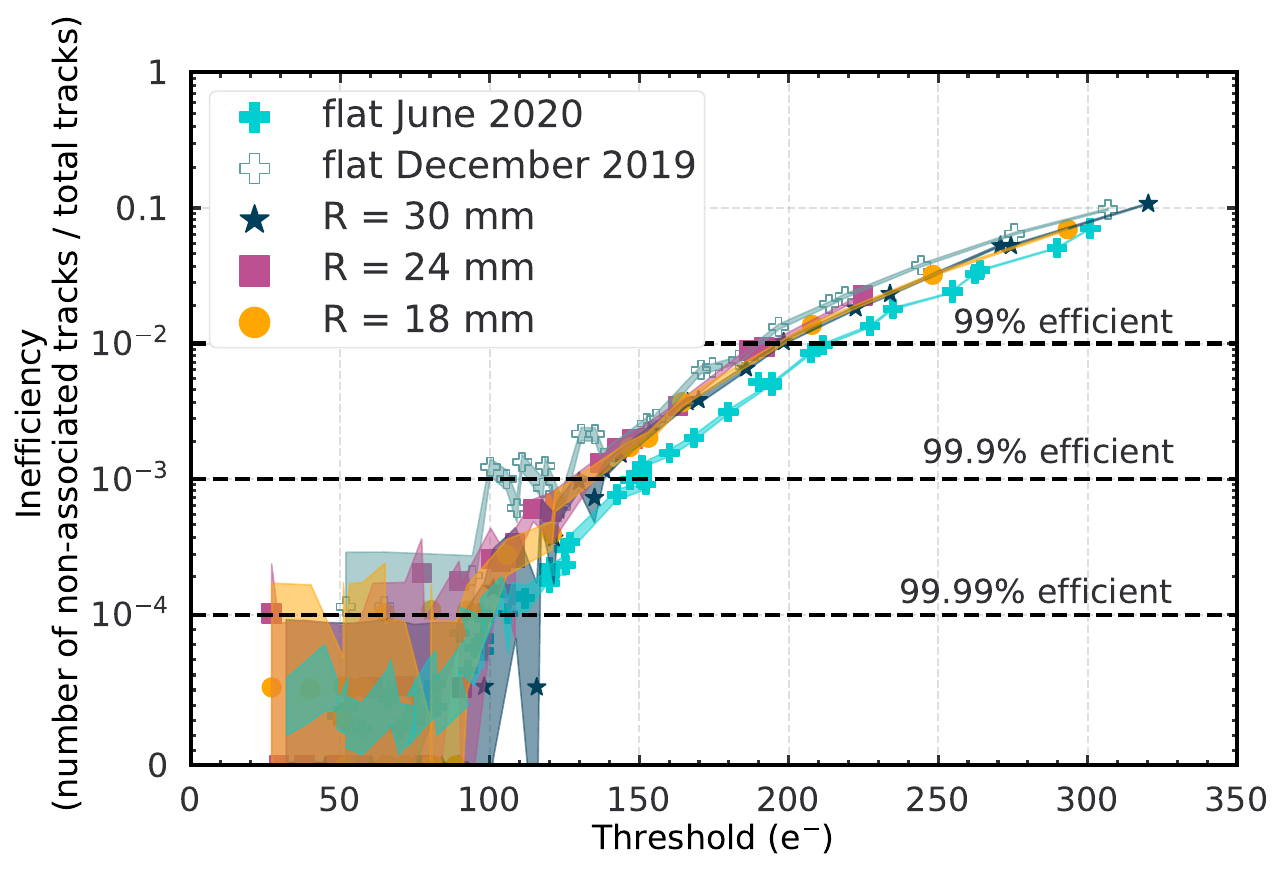}\caption {(Top) Efficiency of bent ALPIDE sensors as a function of the threshold for the three different bending radii compared to flat sensors. (Bottom) Inefficiency of bent ALPIDE sensors as a function of the threshold for the three different bending radii compared to two flat sensors. The nominal operating point for ALPIDE chips is between 100 and 150 electrons corresponding to an average cluster size of about 2.5--3.5. The colored bands indicate the statistical uncertainty.
For illustrative purposes, the y-axis of the inefficiency plot is split in a linear (below $10^{-4}$) and a logarithmic (above $10^{-4}$) part.
  }
    \label{fig:ineff_bent}
\end{figure}

The efficiency $\epsilon$ is defined as the ratio of the number of tracks with an associated cluster on the DUT to the total number of tracks. The inefficieny is defined as $1-\epsilon$. It was determined for each DUT for a range of thresholds. The obtained efficiency and inefficiency for bent sensors compared to flat ones are presented in Fig.~\ref{fig:ineff_bent}. 
For thresholds around 100${e^{-}}$,
an efficiency significantly better than 99\% is achieved, compatible to what was reported previously for a sensor bent along the columns~\cite{ALICEITSproject:2021kcd} and in line with the performance of flat ALPIDE sensors~\cite{freidt_2022}.
A comparison with Fig.~\ref{fig:clsize_vs_thr} shows that in the same threshold range typical average cluster sizes of 2.5--3.5 pixels occur.

The results show no dependence of the efficiency on the bending radii of the sensors, regardless of the operating point.
In principle, smaller radii sensors correspond to on average slightly larger impinging angles of the beam particles. Based on previous measurements~\cite{ALICEITSproject:2021kcd}, one would therefore expect a small increase in efficiency for smaller radii due to enhanced charge sharing in regions of larger angles. However, this effect of the bending radius on the (in)efficiency is not significant in the study presented here due to the limited range of impinging angles 
(\,80$^\circ$ < $\theta$ < 100$^\circ$) covered by the ROI of the DUTs. 

\subsection{Spatial resolution}

\begin{figure}[t]
  \centering

  \includegraphics[width=0.9\textwidth]
  {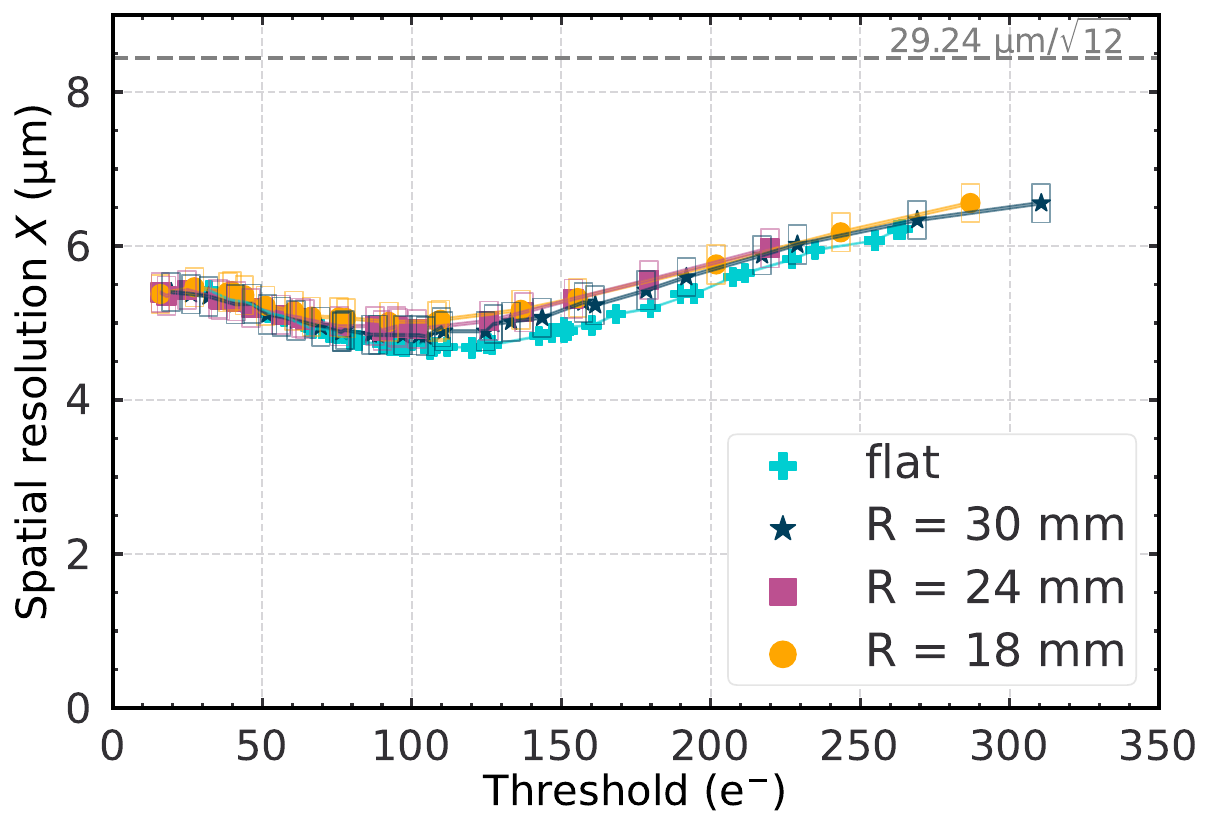}

    \includegraphics[width=0.9\textwidth]
    {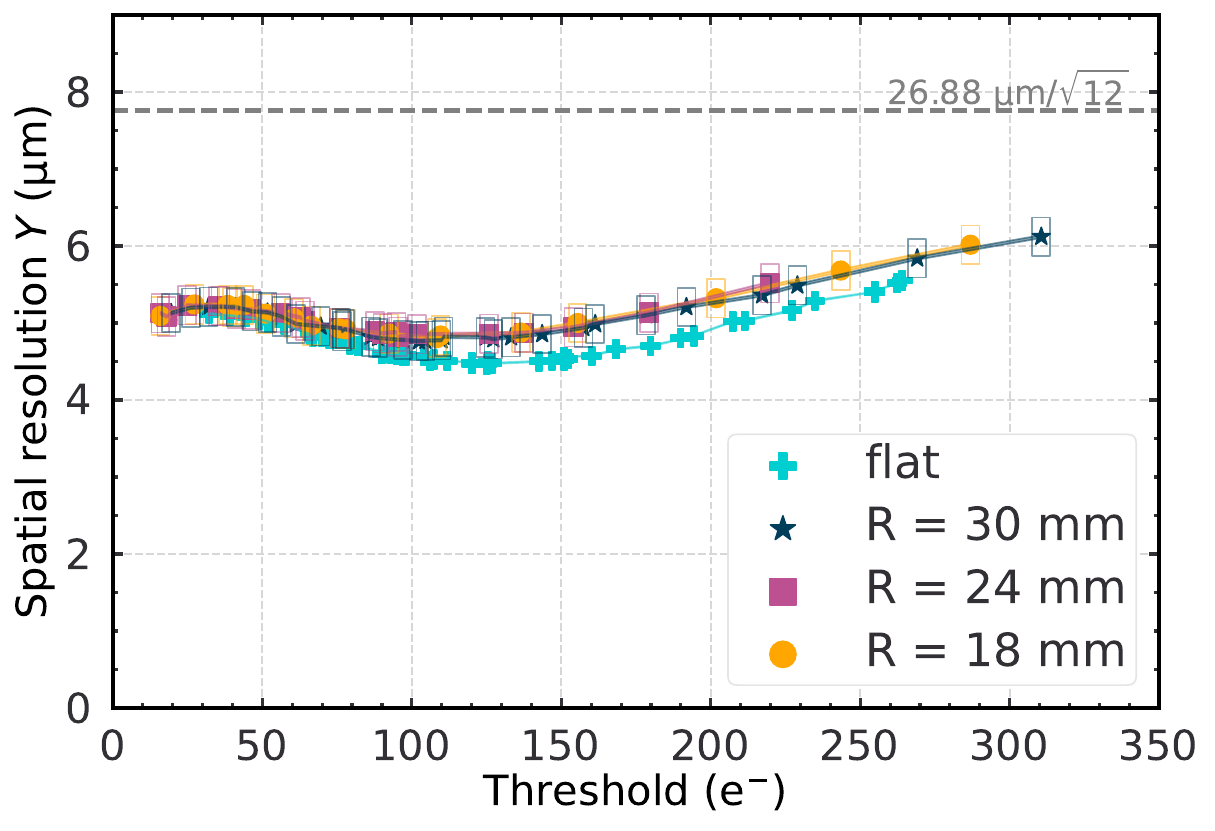}

  \caption {Spatial resolution of ALPIDE chips bent to different radii and flat in the row (top) and in the column (bottom) direction. The box errors indicate the systematic uncertainty on the spatial resolution obtained from the alignment procedure. The intrinsic spatial resolution of the setup has been subtracted.}
  \label{fig:resolution_uITS3chips}
\end{figure}

The space point resolution for the bent ALPIDE chips is determined inside the window (see Fig.~\ref{fig:roi_sketch}) that is not affected by the additional scatterings in the jig material. The track-to-hit residuals are a convolution of the actual space point resolution and the intrinsic uncertainty from the track propagation of the reference tracks.
The latter is determined to be approximately \SI{3.2}{\um} at the position of the DUTs based on a simplified Monte Carlo simulation~\cite{mmagerTelescopeOpt}. This contribution is subtracted in quadrature from the obtained spatial residuals in order to obtain the actual resolution. The systematic uncertainty on the alignment procedure is about~\SI{0.25}{\micro\meter} and was determined by variation of the initial alignment parameters.
The spatial resolution in the bending direction and perpendicular to it are shown in Fig.~\ref{fig:resolution_uITS3chips}. Similarly to the detection efficiency, no significant dependence on the bending radius is observed. As expected from a pixel sensor with digital read-out, an optimal resolution is observed for an average cluster size of 2.5--3.5 pixels corresponding to thresholds in the range of \SIrange[range-phrase = --]{100}{150}{e^{-}}.

\subsection{Grazing beam geometry}\label{sec.:grazing}

With the bent setup described in section~\ref{sec:setup}, other interesting geometries were investigated.
The DUT was rotated by \SI{90}{\degree} from the position shown in Fig.~\ref{fig:setup_sketch} such that the beam can pass the chip surface tangentially along a single row. In the following section, only the \SI{30}{\mm} sensor was studied. For this configuration two track topologies are possible: either the beam particles cross the sensor twice, entering from one side and exiting the other, or they can graze the sensor. In these grazing track topologies, the particles cross tangentially a large length of the active chip volume. Since such setup geometries allow to study simultaneously ordinary single hits and large clusters from grazing collisions within the same chip, they can serve as a precursor for future pixel chamber developments~\cite{MULLIRI2023167724}.

\begin{figure}[t]
  \centering
  \includegraphics[width=0.94\textwidth]{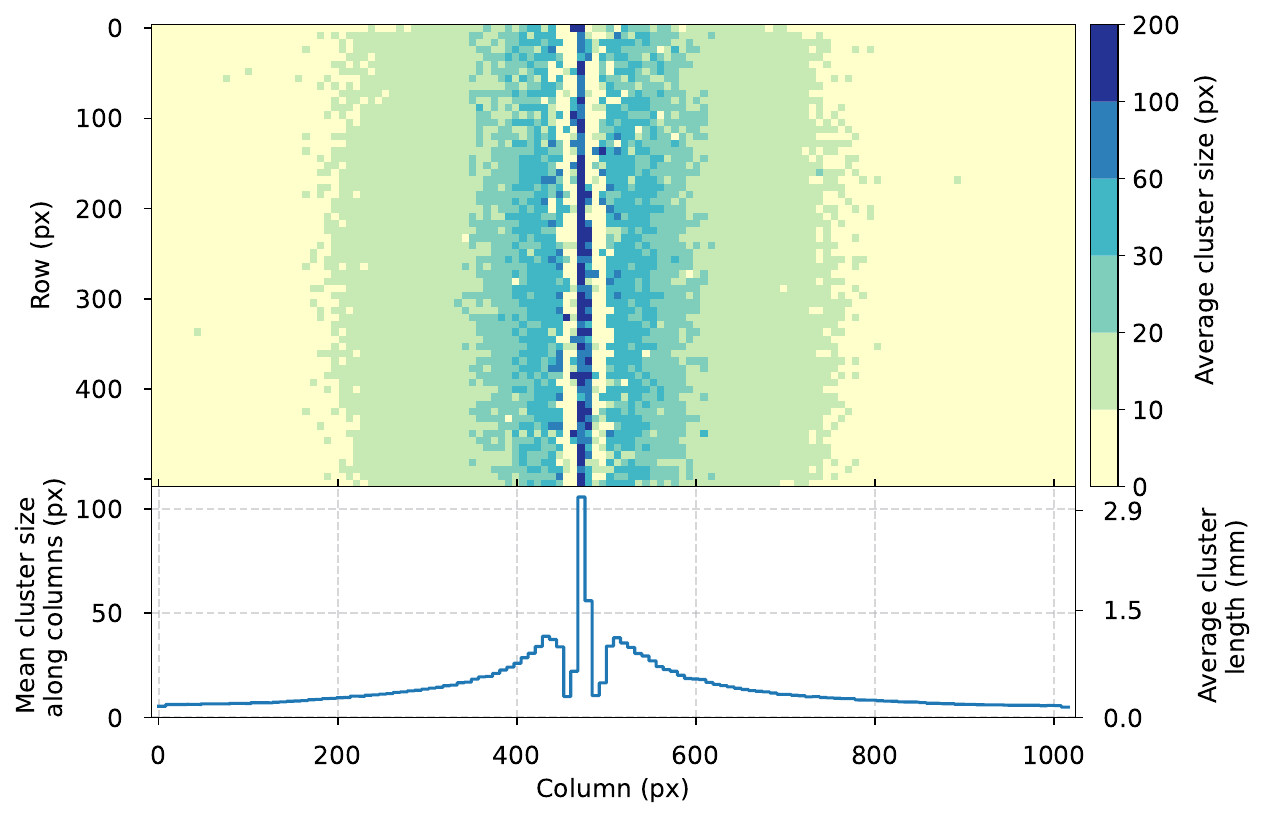}

     \caption {
    (Top) Mean cluster size distribution in the grazing beam configuration. The elongated clusters from grazing collisions are clearly visible in the centre. With the beam crossing the sensor from left to right, towards the edges particles first traverse the sensor on the left side, then enter the sensor again on the right side. (Bottom) Distribution of the centre for these clusters.
    The non-equidistantly spaced color-coded z-axis indicates the average size of the clusters. As can be seen from the projection, very large clusters are found for the grazing incidence. 
    Then transitioning from single continuous clusters in the active volume to two separate clusters produced by one particle, there is a gap where no center of gravity of clusters is reconstructed since pixels will be associated with the large clusters of the grazing incidence. For larger incident angles, smaller and smaller clusters are found. \\
  }
  
  \label{fig:dc_hitmap}
  \label{fig:dc_clustermap}
\end{figure}

The mean cluster size distribution of the chip for this configuration is shown in Fig.~\ref{fig:dc_hitmap}. The elongated clusters are clearly visible in the proximity of the middle of the chip.
Due to the clustering algorithm these elongated clusters are collapsed to their center of gravity.
The length $l$ of these clusters reaches around 100 pixels, i.e.~approximately \SI{3}{\mm}, as expected from simple geometrical considerations ($l=2\sqrt{2br}$) for a tangential track crossing a $b\approx \SIrange[range-phrase = -]{25}{40}{\um}$ thick silicon layer bent to a radius of $r \approx \SI{30}{\mm}$.

\section{Summary}
\label{sec:summary}

The ALICE ITS3 project will be the first application of bent pixel sensors in high energy and nuclear physics experiments.
The results presented in this paper constitute an important milestone towards the development of fully cylindrical silicon tracking detectors based on monolithic active pixel sensors.
They extend and complement previous findings by investigating three different radii, a different bending axis and a setup in which also the periphery is bent.
ALPIDE chips bent to radii of 18, 24 and~\SI{30}{\mm} were proven to preserve performance with respect to their flat state.
The findings demonstrate that an efficiency above 99.9\% is reached in the nominal threshold operating regime. Moreover, a spatial resolution of around \SI{5}{\micro\meter} is achieved. 
Overall, the measurements confirm the absence of effects on the performance of the sensors due to bending, regardless of the bending radius. A similar performance is achieved as with the flat ALPIDE sensors. 
In addition, the peculiar geometry of a grazing incident beam was investigated and mean cluster sizes of more than 100 pixels can be observed.
Bent MAPS allow novel detector concepts and architectures to be realized without any disadvantages compared to traditional geometries.

\newenvironment{acknowledgement}{\relax}{\relax}
\begin{acknowledgement}
\section*{Acknowledgements}
The measurements leading to these results have been performed at the Test Beam Facility at DESY Hamburg (Germany), a member of the Helmholtz Association (HGF). We would like to thank the coordinators at DESY for their valuable support of this testbeam measurement and for the excellent testbeam environment.
F.~Křížek acknowledges the support by the Ministry of Education, Youth and Sports of the Czech Republic project LM2023040.

B.M.~Blidaru and P.~Becht acknowledge support and funding by the HighRR research training group [GRK 2058] and by the German Federal Ministry of Education and Research/BMBF (project number 05H21VHRD1) within the High-D consortium.

This work has been sponsored by the Wolfgang Gentner Programme of the German Federal Ministry of Education and Research (grant no. 13E18CHA).   
\end{acknowledgement}
%
\clearpage
\bibliography{references.bib}
\clearpage
\appendix
\section{Threshold mapping for April 2021 test beam}
\label{sec.:appendixA}

\begin{figure}[htb]
\centering
    \includegraphics[width=0.32\textwidth]{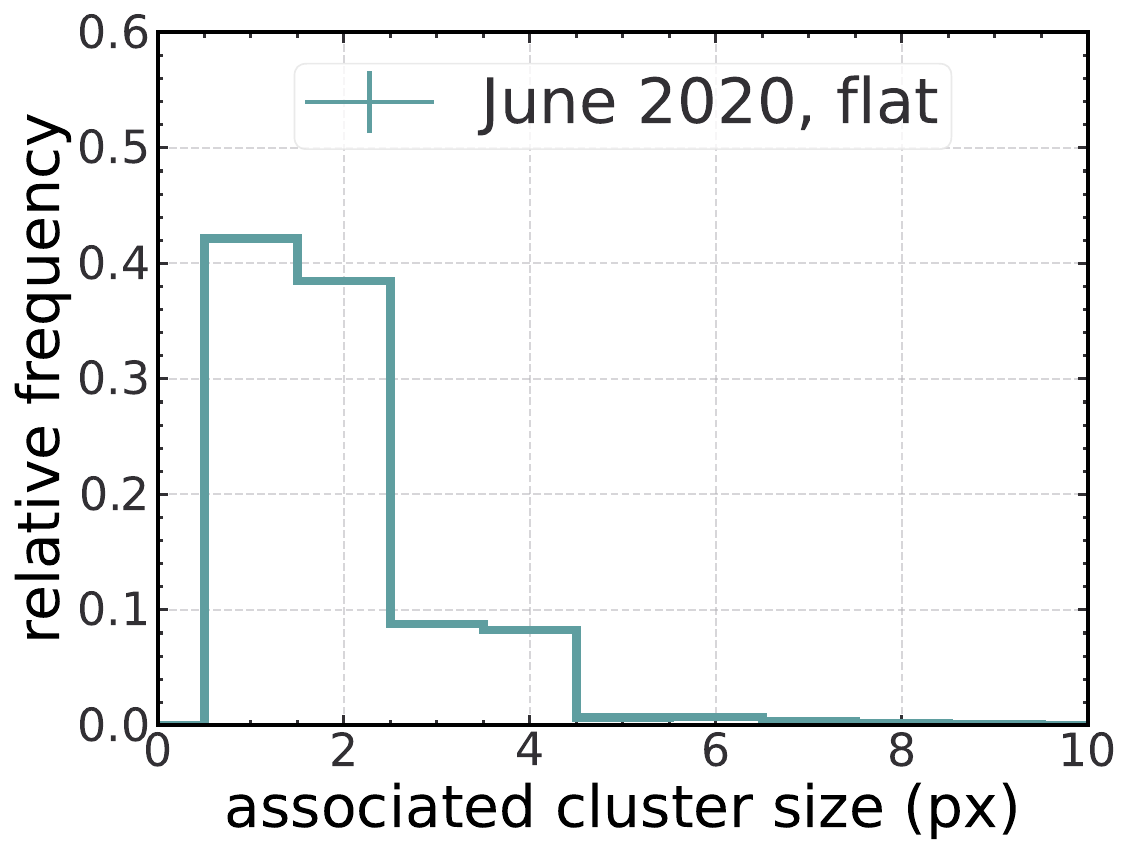}
    \includegraphics[width=0.32\textwidth]{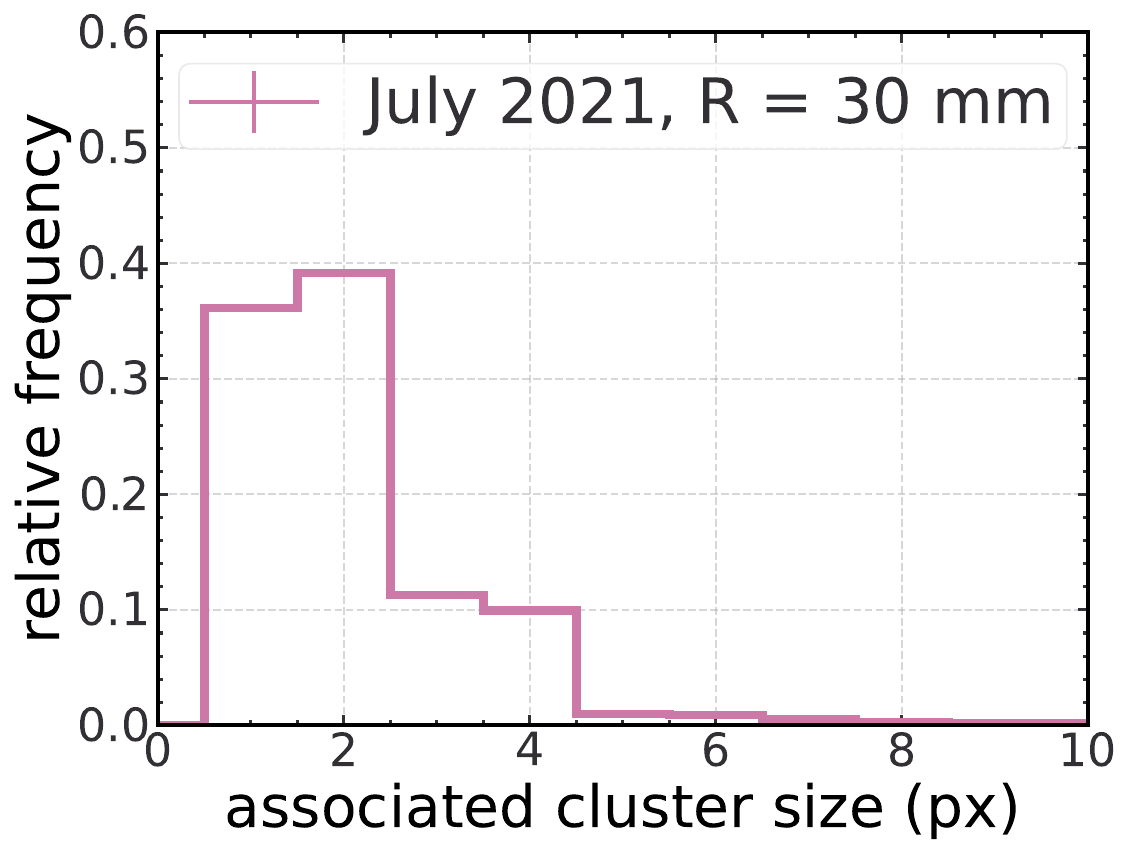}
    \includegraphics[width=0.32\textwidth]{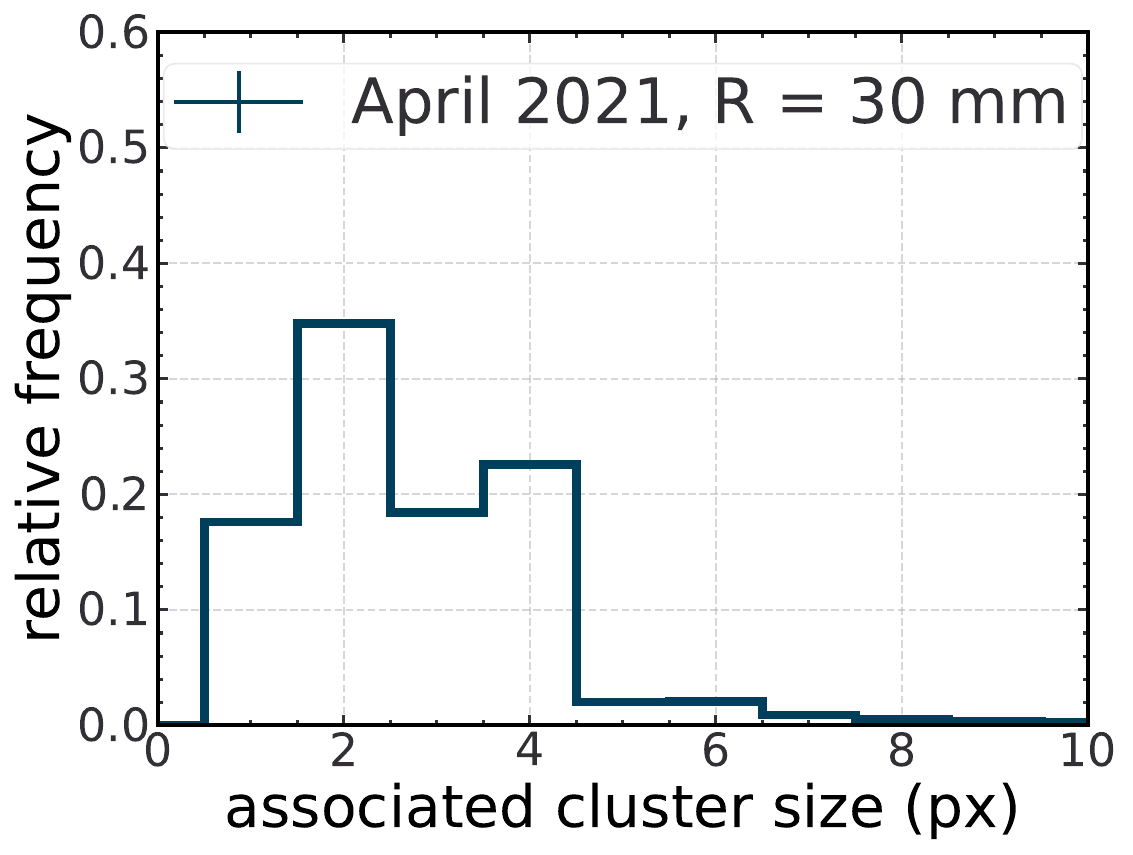}
    
    \includegraphics[width=0.32\textwidth]{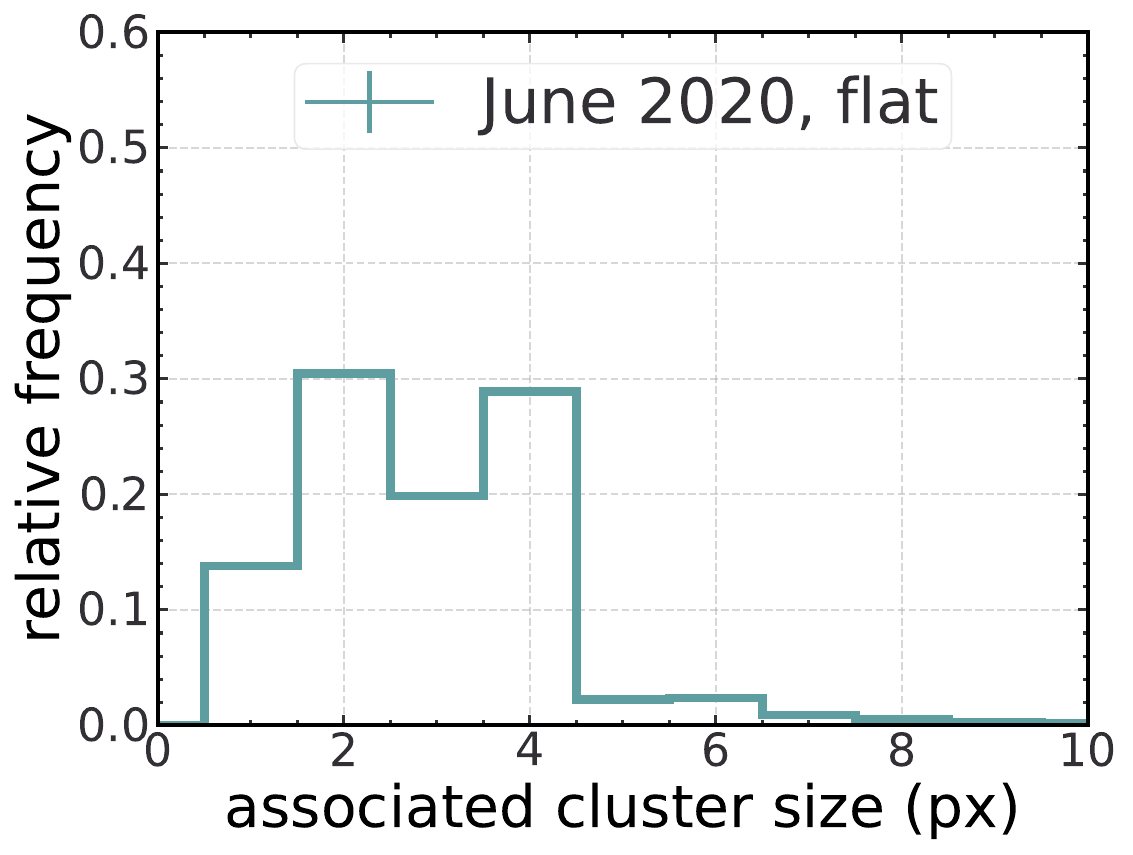}
    \includegraphics[width=0.32\textwidth]{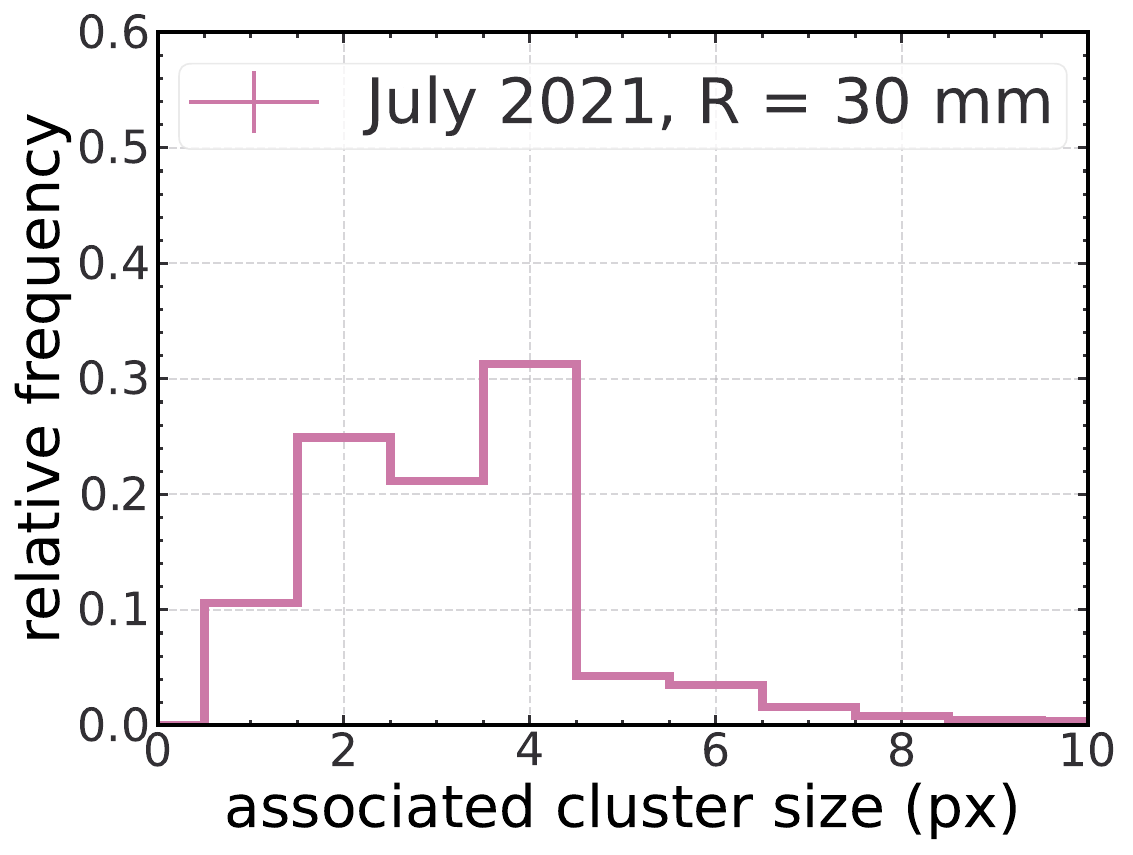}
    \includegraphics[width=0.32\textwidth]{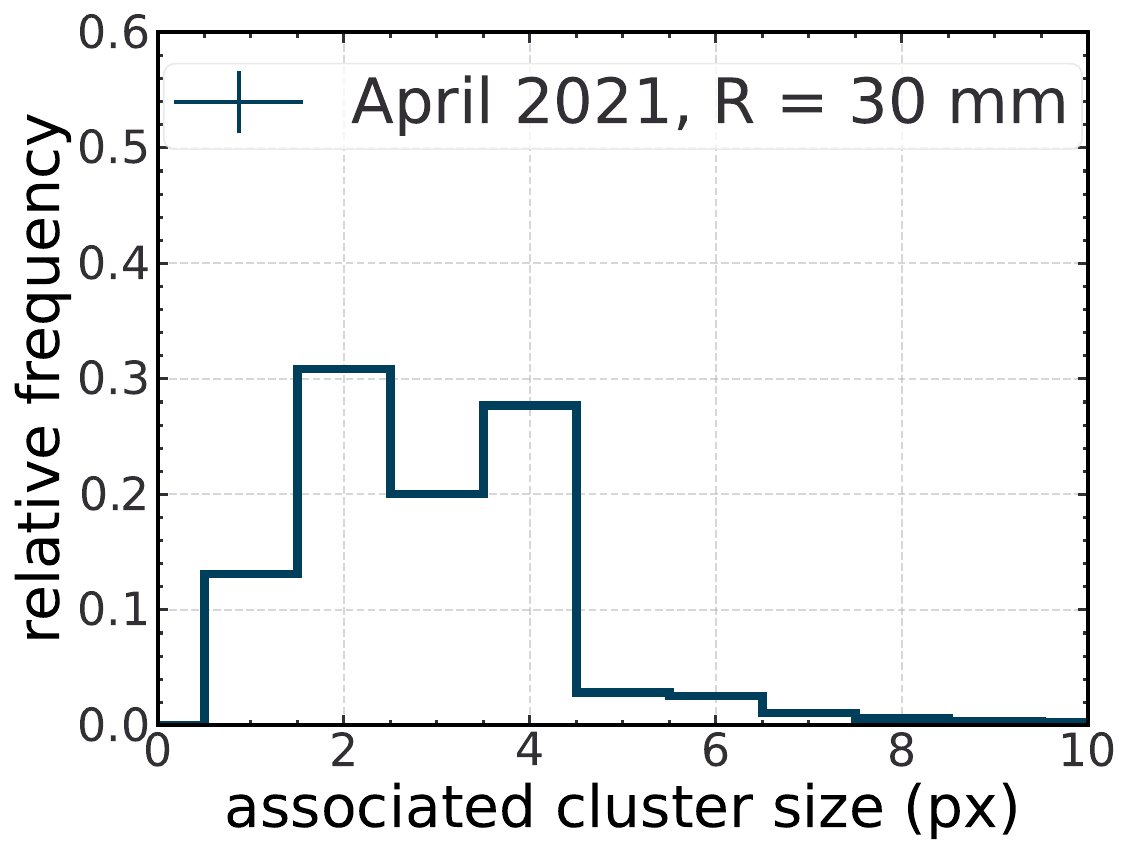}
    \caption{Distribution of the associated cluster size for one flat ALPIDE chip (June 2020 testbeam campaign) compared to the one of bent ALPIDE chips (July 2021 and April 2021 test beam campaign). In the top row shown for a similar threshold of around 220~$e^{-}$ units (see text for details), in the bottom row shown for a similar average associated cluster size of around 3. The thresholds values for the bottom row correspond to 93~$e^{-}$ (June 2020), 118~$e^{-}$ (July 2021), and 88~$e^{-}$ (April 2021).}

    \label{Fig.:ClsSizeDistr}
\end{figure}

The threshold in the ALPIDE sensor is typically measured by a built-in analogue pulsing circuit, injecting discrete charges via a pulsing capacitance to the pixel diodes. These charges correspond to a certain voltage pulse height given by an 8 bit Digital-to-Analog converter (DAC) and can be related to the threshold. One DAC unit corresponds to a charge of 10${e^{-}}$. In contrast to the results from other test beam campaigns presented in this paper, the threshold for the bent ALPIDEs from the DESY test beam in April 2021 could not be reliably determined via internal pulsing.
An example of this problem is illustrated in Fig.~\ref{Fig.:ClsSizeDistr} showing the cluster size distribution of three different chips in one case for similar threshold values (top) and in one case for similar mean associated cluster size (bottom). 
A  distribution of the cluster size is observed for all chips and all threshold settings.
The shape and mean value in case of the April 2021 test beam indicate a different real threshold with respect to the one determined by internal pulsing.
Several investigations to understand the origin of this problem were conducted:
\begin{enumerate}
    
    \item A drop in the supply voltage due to the longer supply lines in the bent setup, namely the FPC and flex cable, could lead to an altered charge injection. This explanation was excluded by measuring the voltages close to the chip on the FPC when pulsing.
    
    \item An influence of additional resistivity and inductance in the bent setup on power draw could bias the injected charge during threshold pulsing. This effect should be larger for a larger number of pixels being pulsed simultaneously and should vary for different decoupling setups. This could be ruled out by varying the number of pixels pulsed from one row to a single pixel in which no significant change was observed. In addition, different decoupling setups (\SI{100}{\nano\farad}, \SI{10}{\micro\farad}, \SI{100}{\micro\farad}) were tried without any measurable effect.
    
    \item An off-center placement of the setup could cause larger incident angles and thus larger cluster sizes due to the larger charge deposit. A detailed investigation of the beam spot position from the first to the last reference plane determined the maximal angular displacement (rotation around $y$-axis) to less than \SI{0.3}{\degree} while angles of more than \SI{10}{\degree} would be required to explain the increased cluster size.

    \item An inter-layer rotation between the \SI{30}{\milli\meter} and \SI{18}{\milli\meter} layer has been observed by studying the correlation between DUTs. This rotation in $x$-axis was determined to be up to \SI{4}{\degree} and does increase the incident angle and therefore the cluster size. This effect was clearly measurable and determined to be 0.15 clusters at the nominal ALPIDE operation conditions. However, this increase corresponds to only about 30\% of the observed effect.
    
    \item Additional rotations of the DUT around the $z$-axis were found to be deviating less than \SI{1.3}{\degree} from the nominal \SI{90}{\degree} rotation. They could be attributed to an insufficiently tightened mounting screw, however with negligible influence on the cluster size.
     
    \item To rule out fundamental differences between bent and flat sensors, $^{55}$Fe source measurements were performed in the laboratory. No difference in the cluster sizes was observed.
    
    \item In literature, different values for the pulsing capacitance were found due to the use of an outdated graphics in several cases. The correct value for the pulsing capacitance could be confirmed to be~\SI{230}{\atto\farad} and is valid for all ALPIDE sensors.
    
    \item Beam contamination with higher ionizing stray particles was ruled out by looking on an event-by-event basis at the average cluster size of the reference planes that was consistent with MIPS.
\end{enumerate}

Since detailed investigation did not reveal the origin of the problem, the most likely explanation was attributed to a detuned potentiometer on the DAQ board which regulates the supply voltage.
All other effects that would lead to either higher cluster sizes or altered thresholds were studied and excluded as described in the list above.
As there is a one to one correspondence of the average cluster size during the run to the actual threshold as shown in Fig.~\ref{fig:clsize_vs_thr}, the threshold for the April 2021 data was thus determined not via the internal pulsing but via this external observable. 
In practice the average cluster size as a function of threshold for the two flat sensors was fitted with a double exponential function as follows:

\begin{footnotesize}
\begin{equation}\label{func:thrfit}
T = \frac{1}{2} \cdot \left(
\cfrac{a}{b + \cfrac{\exp(c \cdot X_{\mathrm{cs}})}{d}} 
+ 
\cfrac{e}{f + \cfrac{\exp(g \cdot X_{\mathrm{cs}})}{h}}
+ 
\cfrac{i}{j + \cfrac{\exp(k \cdot X_{\mathrm{cs}})}{l}} 
+ 
\cfrac{m}{n+ \cfrac{\exp(o \cdot X_{\mathrm{cs}})}{p}}
\right).
\end{equation}
\end{footnotesize}
In this formula $\mathrm{T}$ represents the threshold in electrons, $X_{\mathrm{cs}}$ corresponds to the average cluster size in pixel units, and $a$-$h$ are the phenomenologically determined fit parameters. Using this formula and the obtained fit parameters the true threshold for the April 2021 data was determined based on the average cluster size values.

\end{document}